\documentstyle[preprint,prl,aps,epsfig]{revtex}
\tightenlines

\begin{document}
\preprint{LBL-PUB-42768, January, 1999}
\title{Exclusive Vector Meson Production in 
Relativistic Heavy Ion Collisions} 

\author{Spencer R. Klein and Joakim Nystrand} 

\address{Lawrence Berkeley National Laboratory, Berkeley, CA 94720} 

\break 
\maketitle

\begin{abstract}
Exclusive vector meson production reactions such as Au + Au
$\rightarrow$ Au + Au + V, where $V=\rho, \omega, \phi$ or $J/\psi$
can proceed through photon-Pomeron and photon-meson interactions.
Photons from the electromagnetic field of one nucleus interact
coherently with the other nucleus.  Photonuclear cross sections are
scaled from $\gamma p$ data, and convoluted with the photon spectrum
to find the exclusive rates.

The cross sections at the RHIC and LHC heavy ion colliders are huge,
10\% of the total hadronic cross section at RHIC, and 50\% at LHC.
These accelerators may be useful as vector meson factories. With
iodine beams at RHIC, 640 $\rho$ are produced each second
($10^{10}$/year); with calcium at the LHC the rate is 240 kHz.  The
$\phi$ rates are 39 Hz at RHIC and 15 kHz at LHC, while the $J/\psi$
rate is 0.3 Hz at RHIC and 780 Hz at the LHC.

Because of the coherent couplings, the reactions kinematics are
similar to coherent two-photon interactions; we discuss the interplay
between the two reactions.

\centerline{(Submitted to Physical Review C.)}

\end{abstract}

\narrowtext

\section{Introduction}

The electromagnetic fields of ultrarelativistic heavy ions may be
treated as fields of virtual photons. When two nuclei pass by at small
impact parameters, the photon field of one nucleus can produce a
photonuclear interaction in the other nucleus.  Most studies of
photonuclear reactions in heavy ion collisions have considered hard
interactions, such as heavy quark production\cite{heavyquark} or
photonuclear breakup\cite{breakup}.

We consider here the exclusive reaction $A + A \rightarrow A + A + V$,
where $A$ is a heavy nucleus, and $V$ is a vector meson.  This
reaction can proceed via photon-Pomeron or photon-meson interactions,
where the photon comes from the electromagnetic field of one nucleus,
and the Pomeron or meson couples to the other nucleus.  At
sufficiently high energies, the momentum transfers from the nuclei
are small enough that both couplings are coherent to the entire
nucleus.  Then, the nucleus remains intact, or nearly intact. We
calculate the production rates and rapidity distributions for the
$\rho^0$, $\omega$, $\phi$ and J/$\psi$. 

These reactions will be studied experimentally at the upcoming
Relativistic Heavy Ion Collider (RHIC)\cite{usRHIC} now under
construction at Brookhaven National Laboratory\cite{RHIC}, and the
Large Hadron Collider (LHC)\cite{LHC} being built at CERN.
For later reference, Table \ref{RHIClum} gives the expected
luminosities for various beam species at these two colliders.

\section{Photon flux and photonuclear Cross Sections}

The calculations are simplest in the rest frame of one of the nuclei,
labelled the target nucleus.  The total production is found by taking
each nucleus as the target in turn.  This calculation divides into two
parts: determination of the photonuclear interaction cross section,
and calculation of the effective photon flux from the emitting
nucleus. 

In this frame, the photon energy $k$ is large, and the momentum
transfer squared from the target nucleus is small.  The momentum
transfer $t$ has longitudinal $t_{||}$ and transverse components
$t_\perp$, with $t=t_{||}+t_\perp$.  Since the initial state
longitudinal momentum is small, $t_\perp$ is the final state momentum
transverse to the photon direction.  When $t_\perp=0$,
$|t|=|t_{min}|=\sqrt{M_v^2/4k}$.

The coherent coupling of the photons and Pomerons limits their
virtuality to $1/R_A$, where $R_A$ is the nuclear radius, or about 30
MeV/c for heavy nuclei like gold or lead.  This virtuality is small
enough that it can be neglected here, with the photon effectively real
in the target frame.

The photon flux from a relativistic heavy nucleus is given by the
Weizs\"acker-Williams approach. Here  $\gamma=E_A/m_A\gg 1$, where $E_A$
and $m_A$ are the energy and mass of the nucleus.  The photon flux at
a distance $r$ from the nucleus is
\begin{equation}
{d^3N_\gamma(k,r)\over dk d^2r}= {Z^2\alpha x^2\over\pi^2kr^2} K_1^2(x)
\label{drphotonflux}
\end{equation}
where $k$ is the photon energy, $Z$ is the nuclear charge, $K_1$ a
modified Bessel function with $x=kr/\gamma$.  The target frame
$\gamma$ corresponds to $\gamma=2\Gamma^2-1$, where $\Gamma$ is the
Lorentz boost in the center of mass frame (laboratory frame for
collider geometry).  Here, and throughout the paper, we take
$\hbar=c=1$.  

In an exclusive interaction, there can be no accompanying hadronic
interactions.  In a hard sphere model, this occurs if the impact
parameter $b > 2R_A$, where $R_A=1.2 A^{1/3}$\ fm.  We use a
more accurate approach, calculating, the probability of one or more
hadronic interactions as a function of impact parameter.  The nuclear
density for a nucleus $A$ a distance $s$ from its center is modelled
with a Woods- Saxon distribution for symmetric nuclei
\begin{equation}
\rho_A(s)= { \rho_0 \over 1 + exp[(s-R_{WS})/d]}
\end{equation}
where the radius $R_{WS}$ and skin depth $d$ are based on fits to
electron scattering data\cite{elects} and $\rho^0$ is fixed by the
normalization.  This electron scattering data probes the proton
distribution in the nucleus.  If the neutron are differently
distributed from the protons, this could affect the cross section..

The interaction probability at a given impact parameter is
related to the overlap function
\begin{equation}
T_{AA}(|\vec{b}|) = \int d^2\vec{r} T_A(\vec{r}) T_A(\vec{r}-\vec{b})
\end{equation}
where $\vec{r}$ and $\vec{b}$ are 2-dimensional vectors perpendicular
to the direction of travel $z$. The nuclear thickness function is
\begin{equation}
T_A(\vec{r})=\int dz \rho_A(\sqrt{|\vec{r}|^2+z^2})
\end{equation}
The number of nucleon-nucleon collisions follows a Poisson
distribution with mean $T_{AA}(b)\sigma_{NN}$, with $\sigma_{NN}$ the
total nucleon-nucleon interaction cross section. The probability of
having no hadronic interactions is then
\begin{equation}
P_{0H}(b)=\exp(-T_{AA}(b)\sigma_{NN}).
\label{eqhint}
\end{equation}
We use $\sigma_{NN}= 52$ mb at a beam energy of 100
GeV, rising to 88 mb at 2.9 TeV\cite{PDG}.

We assume that the range of the Pomeron or meson field is much smaller
than the size of the nucleus \cite{muller}.  For $b<2 R_A$, a form
factor can be required to accurately model the electromagnetic fields
inside the emitting nucleus; we neglect this tiny correction.

\begin{figure}
\setlength{\epsfxsize=0.7\textwidth}
\setlength{\epsfysize=0.7\textheight}
\centerline{\epsffile{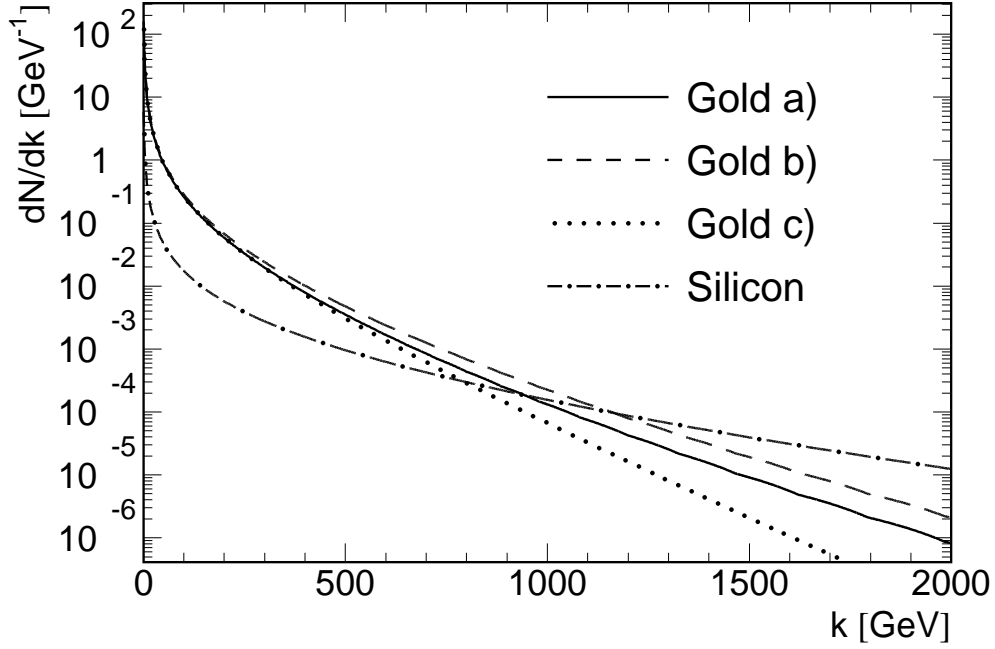}}
\caption[]{The photon flux for gold beams using (a) the hadronic
interaction probability described in the text, (b) the hard sphere
approximation, (c), the photon flux for $b>2R_A$,
Eq. (\ref{sphotonflux}), and for silicon beams with the same approach
as in (a).  Both are at RHIC energies, with $k$ in the target frame..}
\label{fphotonflux}
\end{figure}

The total photon flux is given by Eq. (\ref{drphotonflux}) integrated over
all $r$, modulated by the noninteraction probability.  
\begin{equation}
{dN_\gamma(k) \over dk} =  
\int_0^{\infty} 2\pi bdb P_{0H}(b)
\int_0^R  {rdr \over \pi R_A^2} 
\int_0^{2\pi} d\phi 
\ \ {d^3N_\gamma(k,b+r\cos(\phi))\over dkd^2r}
\label{ephotonflux}
\end{equation}
The $b$ integral runs over impact parameter and $r$ and $\phi$ over
the surface of the target nucleus, for a given $b$.  This process
averages the field strength over the nucleus, neglecting the variation
with $r$. Since the vector meson production is coherent over the
entire nucleus, this averaging is appropriate. The integral is
evaluated numerically; the result is shown in Fig.~\ref{fphotonflux}.
This flux is approximately equal to the photon flux in the region
$r>2R_A$, which can be found analytically:
\begin{equation}
{dN_\gamma(k)\over dk} = {2Z^2\alpha\over\pi k} \bigg(XK_0(X) K_1(X) -
{X^2\over 2}(K_1^2(X) - K_0^2(X))\bigg)
\label{sphotonflux}
\end{equation}
where $X=2R_A k/\gamma$.  This approximation is shown by the
dotted line in Fig.~\ref{fphotonflux}. 

The $\gamma A\rightarrow V A$ cross sections are found with a Glauber
calculation, using the observed $\gamma p\rightarrow Vp$ cross
sections as input.  Data on $\gamma p\rightarrow Vp$ covers a wide
energy range, from low energy fixed target photon beams\cite{review}
to $ep$ collisions at HERA\cite{nicolo}.  These cross sections may be
parameterized as
\begin{equation}
{d\sigma(\gamma p\rightarrow V p)\over dt}\bigg|_{t=0} 
= b_V (X W^\epsilon + Y W^{-\eta})
\end{equation}
where $W$ is the center of mass energy, and Table~\ref{HERAsigma}
gives the constants $b_V$, $X$, $Y$, $\epsilon$ and $\eta$, which are
determined from fits to data\cite{sigmacon}.  The $X$, $\epsilon$ term
represents the Pomeron exchange portion of the cross section.  It
rises slowly with energy ($\epsilon\sim 0.22$).  This slow rise has
been attributed to the increasing size and density of the proton; the
increasing $W$ probes smaller proton momentum fractions $x$.  The $Y$,
$\eta$ term, is for meson exchange, primarily the $f_0$
\cite{leith}. The meson exchange term falls rapidly as $W$ rises.  For
the $\phi$ and $J/\psi$, meson exchange is heavily suppressed, and the
reaction occurs only through Pomeron exchange.  For the $J/\psi$,
$\epsilon\gg 0.22$; the rapid rise has been ascribed to either a
breakdown of the soft Pomeron model or to threshold
effects\cite{nicolo}.  Because of this behavior, the $J/\psi$
calculations must be treated with caution, especially at center of
mass energies beyond that probed by HERA.

To relate this to nuclei, we make use of the optical theorem
and an Eikonalization technique.
Following vector dominance\cite{review},
\begin{equation}
{d\sigma(\gamma p\rightarrow Vp) \over dt} \bigg|_{t=0} = {4\pi\alpha
\over f_v^2} \ \ {d\sigma (Vp\rightarrow Vp)\over dt} \bigg |_{t=0}
\end{equation}
with a similar relationship true for nuclei.  Here, t is the squared
4-momentum transfer between the proton and vector meson, $\alpha$ is
the electromagnetic coupling constant, $e^2/\hbar c$, and $f_v$ is the
vector meson-photon coupling,
\begin{equation}
{f_v^2 \over 4\pi} =  { M_v\alpha^2 \over 3\Gamma_{V\rightarrow e^+e^-}}
\label{fvs}
\end{equation}
with $M_v$ the vector meson mass and $\Gamma_{V\rightarrow e^+e^-}$
the leptonic decay partial width.  However, measurements of $f_v$ from
leptonic decay widths disagree with photoproduction data\cite{pautz}.
A correction is required to account for non-diagonal coupling through
higher mass vector mesons, as implemented in the generalized vector
dominance model (GVDM).  For the $\rho$, GVDM increases $f_v^2$ by
20\%\cite{pautz}. Because $f_v^2$ is used twice, the error largely
cancels here.  Values for $f_v^2/4\pi$ are given in
Table~\ref{HERAsigma}.

\begin{figure}
\setlength{\epsfxsize=0.8\textwidth}
\setlength{\epsfysize=0.8\textheight}
\centerline{\epsffile{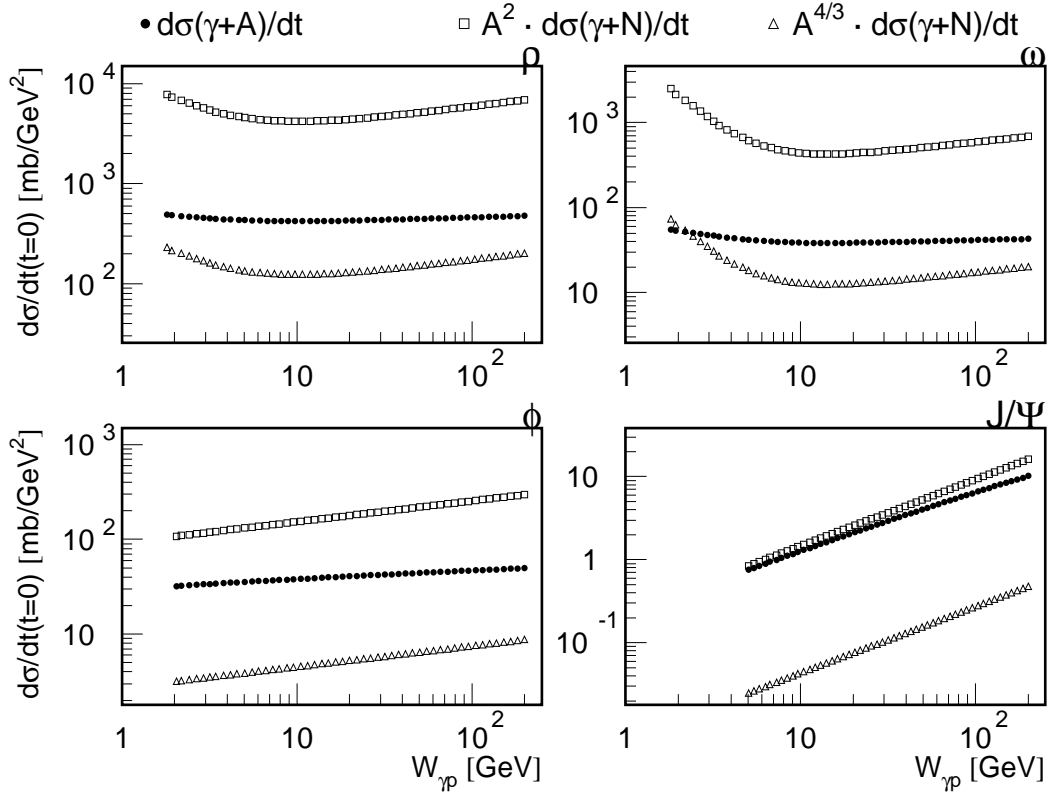}}
\caption[]{$d\sigma/dt|_{t=0}$ for coherent $\rho$, $\omega$, $\phi$
and $J/\psi$ production on a gold target. The solid circles are the
Glauber calculation, while the open squares and triangles show the
$A^2$ weak absorption and $A^{4/3}$ black disk scaling results
respectively.  The $J/\psi$ is in the weak interaction regime, while
the lighter mesons are closer to the black disk.  The Glauber
calculation removes much of the $\gamma p$ cross section energy
dependence visible with the simple scaling.}
\label{dsdtgamma}
\end{figure}

Using the optical theorem, the total cross section is
\begin{equation}
\sigma_{tot}^2(Vp) = 16\pi {d\sigma(Vp\rightarrow Vp)\over dt}
\bigg|_{t=0}.
\end{equation}
At $W=10$ GeV, we find 24 mb, 26 mb, 12 mb, and 1.0 mb for the $\rho$,
$\omega$, $\phi$, and $J/\psi$ total cross sections respectively.
These values are consistent with theoretical and experimental
expectations\cite{review}. The scattering cross section from heavy
nuclei can be found by a Glauber calculation
\begin{equation}
\sigma_{tot}(VA) = \int d^2 \vec{r} \bigg(1 -
e^{-\sigma_{tot}(Vp) T_{AA}(\vec{r})}\bigg).
\end{equation}
The optical theorem for nucleus $A$ and GVDM are then used to find
\begin{equation}
{d\sigma(\gamma A\rightarrow VA)\over dt} \bigg|_{t=0} =
{\alpha\sigma_{tot}^2(VA) \over 4f_v^2}
\end{equation}
These $d\sigma/dt|_{t=0}$ are shown in Fig.~\ref{dsdtgamma} for gold
nuclei. Also shown are the results of scaling the $\gamma p$ cross
sections by $A^2$ and $A^{4/3}$\cite{usRHIC}. The $\rho$, $\omega$,
and $\phi$ results vary only slowly with energy.  The minimum around
photon-proton center of mass energy $W_{\gamma p}=10$ GeV for the
$\rho$ and $\omega$ marks the transition from meson dominated to
Pomeron dominated reactions.  $\phi$ production, which is mediated
only by Pomerons, shows a monotonic energy dependence.

For the light mesons, $d\sigma/dt|_{t=0}$ is almost independent of
energy, flatter than the corresponding proton functions, which appears
unchanged in the $A^2$ and $A^{4/3}$ scaling curves.  In contrast,
$d\sigma/dt|_{t=0}$ for the $J/\psi$ rises rapidly with energy.  Part
of the difference is the different $\gamma p$ cross section energy
dependence, and part is the smaller $J/\psi$ interaction cross
section.  Because $\sigma_{\rho N}$ is large, $\rho$ interact near the
nuclear surface, and $\sigma_{\rho A}$ is the geometric cross section
$\pi R_A^2$ which is almost independent of $k$.  However,
$\sigma_{J/\psi N}$ is much smaller, and the $J/\psi$ interactions
occur throughout the nucleus.  As long as $\sigma_{J/\psi N} R_A
\rho_0 \ll 1 $, the rising $\sigma_{J/\psi N}$ will reflect itself in
$\sigma_{J/\psi A}$.  These cases correspond to the black disk and
weak absorption limits respectively.

The HERA results which provide the high energy $\gamma p$ input data
extend up to about $W_{\gamma p}$ = 180 GeV ($k=16$ TeV in the target
frame), with higher energy points extrapolations.  Although the HERA
data covers the entire RHIC energy range, the extrapolations are
required for LHC.  The uncertainty is especially problematic for the
$J/\psi$, where $d\sigma/dt|_{t=0}$ must eventually flatten out, to
avoid dominating the total photonuclear cross section.

The total cross section depends on the slope of $d\sigma/dt$.  For
proton targets, the cross section can be parameterized as $d\sigma/dt=
A_v exp(-bt+c|t|^2)$\cite{ZEUSrho}.  The slope $b$, and derivative $c$
represent the sum of the proton size and the range of the interaction.
Here, for simplicity we take $c=0$ with little loss of accuracy.
Nuclei are much larger than protons, so $b$ is dominated by the
nuclear size, with $b\sim R^2$ and $d\sigma/dt$ is dominated by the
nuclear form factor. This is important because experimental study of
these exclusive reactions depends on their small
$t_\perp$\cite{goldflash} which is determined by the form factor.

\begin{figure}
\setlength{\epsfxsize=0.7\textwidth}
\setlength{\epsfysize=0.7\textheight}
\centerline{\epsffile{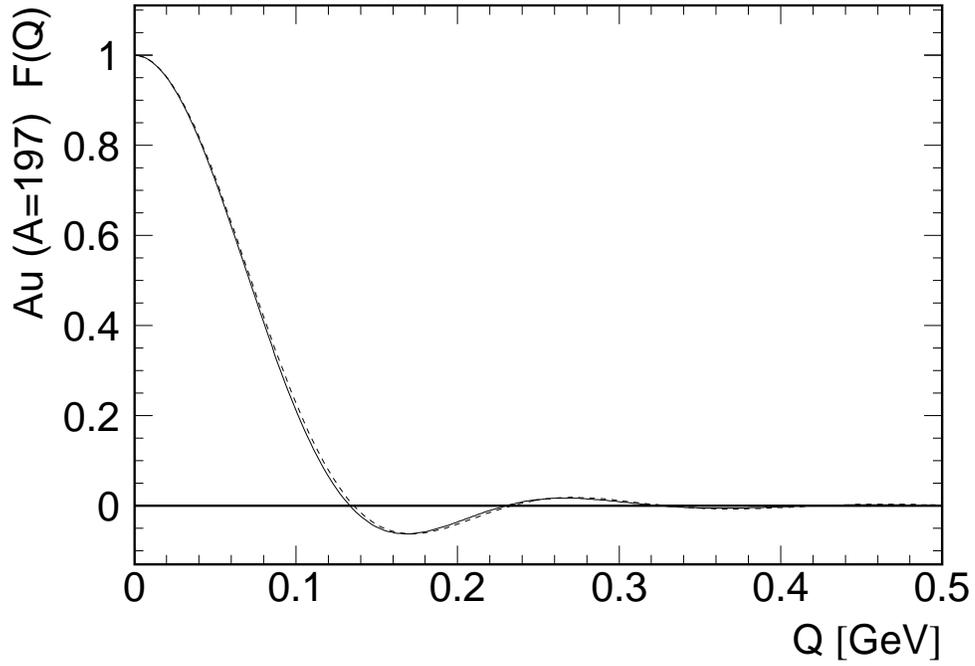}}
\caption[]{The electromagnetic form factor for gold. The solid line is
the exact result from Fourier transformation of a Woods-Saxon
potential, and the dotted line from Eq. (\ref{eqff}).}
\label{fformfactor}
\end{figure}

The form factor is the Fourier transform of the nuclear density
profile.  Unfortunately, the Woods-Saxon distribution does not have an
analytic form factor.  Instead, for simplicity we approximate the
Woods-Saxon distribution as a hard sphere, with radius $R_A$,
convoluted with a Yukawa potential with range $r=0.7$ fm\cite{nix}.
The Fourier transform of this convolution is the product of the two
individual transforms:
\begin{equation}
F(q=\sqrt{|t|}) = {4\pi\rho_0\over Aq^3}
\bigg[\sin(qR_a)-qR_a\cos(qR_A)\bigg] \ \ \bigg[{1\over1+a^2q^2}\bigg]
\label{eqff}
\end{equation}
Figure~\ref{fformfactor} compares Eq. (\ref{eqff}) with the numerical
transform of the Woods-Saxon distribution for gold.  The agreement is
excellent. 

In addition to coherent production, there is also incoherent vector
meson production, where the photon interacts with a single nucleon in
the target.  Incoherent interactions have a larger average $t$, and
occur at a somewhat lower rate.  They will not be further considered
here, but they will be an essential component of any experimental
analysis.

\begin{figure}
\setlength{\epsfxsize=0.7\textwidth}
\setlength{\epsfysize=0.7\textheight} 
\centerline{\epsffile{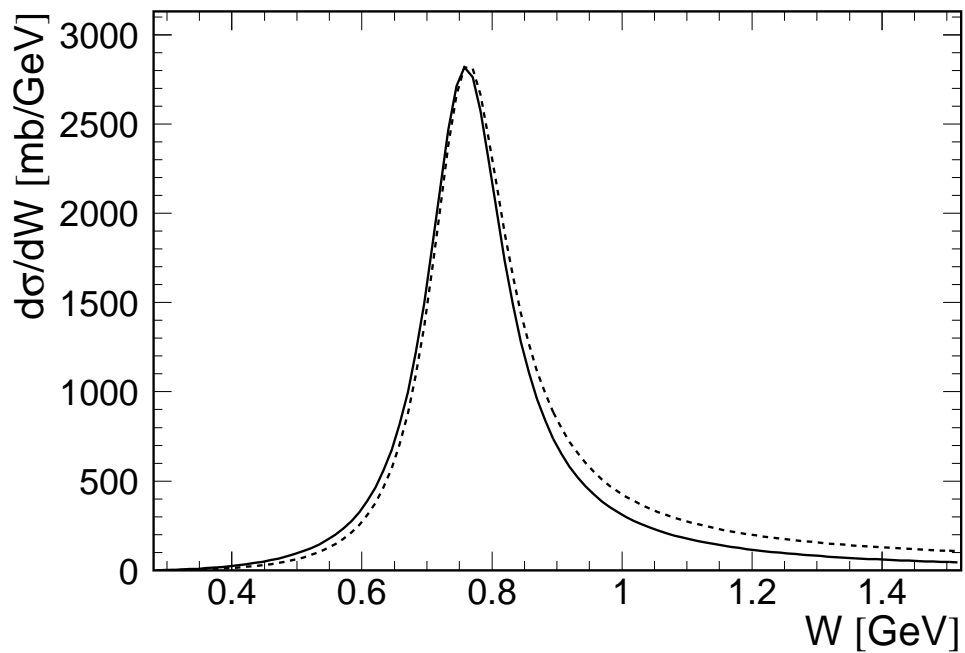}}
\caption[]{Predicted $M_{\pi\pi}$ spectrum for exclusive $\rho$
photoproduction (solid line), compared with the input Breit-Wigner
spectrum (dashed line).}
\label{rhospectrum}
\end{figure}

The photonuclear cross section is
\begin{equation}
\sigma(\gamma A\rightarrow VA) = {d\sigma(\gamma A\rightarrow
VA)\over dt}\bigg|_{t=0} \int_{t_{min}}^\infty dt |F(t)|^2
\end{equation}
For narrow resonances, $t_{min}=(M_v^2/4k)^2$. Because of its width, the
$\rho$ is more complicated, and the cross section must be calculated
using a Breit-Wigner resonance:
\begin{equation}
{d\sigma\over dM_{\pi\pi}} = 
{2\over\pi}\ {\sigma_0 \Gamma_\rho M_\rho M_{\pi\pi} \over
(M_{\pi\pi}^2-M_\rho^2)^2 + \Gamma_\rho^2M_\rho^2}. 
\end{equation}
Here, $M_\rho$ is the pole position of the resonance, $\sigma_0$ the
total cross section (neglecting phase space corrections) and
$M_{\pi\pi}$ the final state invariant mass.  The Breit-Wigner width
$\Gamma_\rho$ includes the phase space correction, with $\Gamma_\rho =
\Gamma_0 (p_\pi/p_0)^3 (M_\rho/M_{\pi\pi})$ \cite{oldjackson} where
$p_\pi$ is the decay pion momentum in the $\rho$ rest frame, with
$p_0$= 358 MeV/c the pion momentum for $M_{\pi\pi}=M_\rho$ and
$\Gamma_0$ is the pole $\rho$ width.  The observed resonance shape is
the convolution of this Breit-Wigner with the photon spectrum.  Figure
\ref{rhospectrum} compares the resulting spectrum with the input
Breit-Wigner.  With the Breit-Wigner, the cross section is about 5\%
lower than for a narrow resonance with the same coupling.  The phase
space correction naturally cuts off the $M_{\pi\pi}$ spectrum at
$2m_\pi$; we add an upper limit $M_{\pi\pi} < M_\rho+5\Gamma_0$,
matching the HERA analysis from which the $\gamma p$ parameters are
extracted\cite{ZEUSrho}.  The correction for higher masses is a few
percent.

Most studies of $\gamma p\rightarrow \rho p$ have modelled the
exclusive $\pi^+\pi^-$ spectrum with a Breit-Wigner $\rho$ plus a
non-resonant $\pi^+\pi^-$ term, and an interference term\cite{soding}:
\begin{equation}
{d\sigma \over dM_{\pi\pi}} =
\bigg| { A \sqrt{M_{\pi\pi}M_\rho\Gamma_\rho} \over
M_{\pi\pi}^2 - M_\rho^2 + i M_\rho\Gamma_\rho} + B \bigg|^2.
\end{equation}
The ZEUS collaboration found $A=-2.75\pm 0.04$ $\mu$b$^{1/2}$ and
$B=1.84\pm0.06$ ($\mu$b/GeV)$^{1/2}$\cite{ZEUSrho}.  With the
interference term, $d\sigma/dM_{\pi\pi}$ falls rapidly at large $W$,
and a correction for $M_{\pi\pi}>M_\rho+5\Gamma_0$ is negligible.

As nuclear size increases, the non-resonant term becomes less
important.  At lower energies, for heavy nuclei it is at most a few
percent.  For large nuclei, the $k$ dependence of $B$ is unknwon, so
we neglect non-resonant production.  For the nuclei considered, this
should be less than a 10\% effect on the $\pi^+\pi^-$ rate.

\begin{figure}
\setlength{\epsfxsize=0.7\textwidth}
\setlength{\epsfysize=0.7\textheight}
\centerline{\epsffile{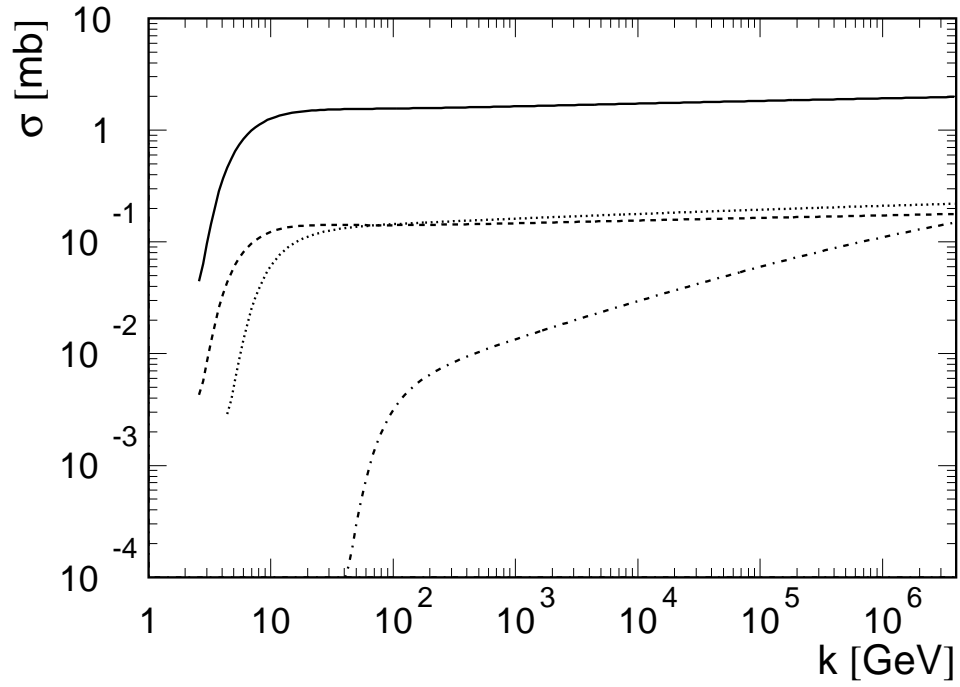}}
\caption[]{Photonuclear cross sections $\sigma(\gamma A\rightarrow VA)$
for $\rho^0$ (solid line), $\omega$ (dashed line), $\phi$ (dotted
line) and $J/\psi$ (dot-dashed line) coherent production on a gold
target.  }
\label{sigmagamma}
\end{figure}

The photonuclear cross sections are shown in Figure~\ref{sigmagamma}.
The total cross sections follow a similar pattern to
$d\sigma/dt\big|_{t=0}$. At low energies, the nuclear form factor
intrudes, and eliminates coherent production, with the cutoff energy
depending on the nuclear radius and the final state mass. The $\omega$
and $\phi$ curves cross because the $\omega$ includes a meson
contribution that decreases with increasing energy, while the $\phi$
does not and because $\sigma_{\phi N} < \sigma_{\omega N}$, with the
exact relationship varying slightly with energy.

Coherent photonuclear interactions are not a new concept.  Several groups
studied them experimentally and theoretically in the 1960s and 1970's,
using real photon beams.  More recent experiments, using virtual photons
from muon scattering are less comparable because of the significant
(compared with $1/R_A$) photon virtuality.

Most of the earlier calculations were used to extract parameters such
as $\sigma_{VN}$ and $R_A$ from data. Our calculations are broadly
similar. However, most of the earlier calculations used a disk model
of the nucleus, with $t_\perp$ determined by a 2-dimensional Fourier
transform of a black disk, and $t_{||}$ contributing only to a phase
shift. Many of these calculations also included a real part for the
forward scattering amplitude.  For the $\rho$, the real part is
20-30\% of the imaginary part, producing a 5\% correction to the total
amplitude\cite{review}.  For other mesons, the phases should be
comparable (when meson exchange is present) or smaller, for
Pomeron-only interactions.  In all cases, the correction should
decrease slowly at higher energies.

The experimental results can also test our calculations. Because of
the complications due to differing (often poorly described) treatments
of backgrounds, $\rho$ width and the nonresonant $\pi^+\pi^-$, we
limit our comparisons to narrow resonances.  Even for narrow
resonances, there are complications. Few experiments cover the
complete $t$ range; extrapolations are required to find the cross
sections.  We compare results using $d\sigma/dt|_{t=0}$, because less
extrapolation is required.

An experiment at Cornell studied $\omega$ production from photons with
average energy 8.2 GeV striking a copper target, and found
$d\sigma/dt|_{t=0}=9.6\pm 1.2$ mb/GeV$^2$\cite{data1}.  This compares
well with our calculated $d\sigma/dt|_{t=0}=9.5$ mb/GeV$^2$ for that
system.  Another group studied $\phi$ production from 8.3 GeV photons
incident on copper and lead targets, and found $d\sigma/dt|_{t=0}=
4.1\pm 0.7$ and $19\pm 3$ mb/GeV$^2$ respectively\cite{data2}.  We
predict higher cross sections, $d\sigma/dt|_{t=0}=5.4$ and 38
mb/GeV$^2$ respectively.  However, with the values of $f_\phi$ and
$\sigma_{\phi N}$ used in the original analysis, we find
$d\sigma/dt|_{t=0}=3.9$ and 26 mb/GeV$^2$ respectively, in agreement
with the data.  The discrepancy may stem from the large
correction(60\% for copper, 140\% for lead) to cover the full angular
acceptance.  This correction is based on the optical model, and so
depends on $f_\phi$ and $\sigma_{\phi N}$.  Still, these comparisons
provide a useful check of our methods.

\section{heavy ion cross sections}

By integrating the photonuclear cross section over the photon
spectrum, the total cross section is found:
\begin{equation}
\sigma(AA\rightarrow AAV) = \int dk {dN_\gamma(k)\over dk}
\sigma(\gamma A\rightarrow VA)=
\int_0^\infty dk {dN_\gamma(k)\over dk} \int_{t_{min}}^\infty
dt {d\sigma(\gamma A\rightarrow VA)\over dt}\bigg|_{t=0} 
|F(t)|^2
\end{equation}
where the photon flux is from Eq. (\ref{ephotonflux}).  The total
cross sections for vector meson production at RHIC and LHC for a
variety of beam species are shown in Table~\ref{sigmavm}.  The
corresponding production rates are shown in Table~\ref{ratevm}.  These
rates are very high, with exclusive meson cross sections about 10\% of
the total hadronic cross section for gold at RHIC, and 50\% for lead
at LHC.  If the hadronic interaction model, Eq. (\ref{eqhint}), is
replaced with a hard sphere cutoff, ($b>2R_A$), the rates rise about
5\%.

Previously, we considered a model where $\gamma A$ cross sections were
scaled from $\gamma p$ treating the nucleus as a black disk, with
$\sigma\sim A^{2/3}$.  In that model, the photonuclear cross sections
had a different $k$ dependence, shown by the open triangles in
Fig.~\ref{dsdtgamma}.  This scaling predicted cross sections much
lower than are found here\cite{sn347}.  Most other studies of
photon-Pomeron interactions have neglected the coherent
Pomeron-nucleus coupling.  Baur, Hencken and Trautman found
$\sigma(Aup\rightarrow AupV)$=3.5 mb using the Weizs\"acker-Williams
photon flux for gold incident on a free proton target at
RHIC\cite{baurrev}.

\begin{figure}
\setlength{\epsfxsize=0.7\textwidth}
\setlength{\epsfysize=0.7\textheight}
\centerline{\epsffile{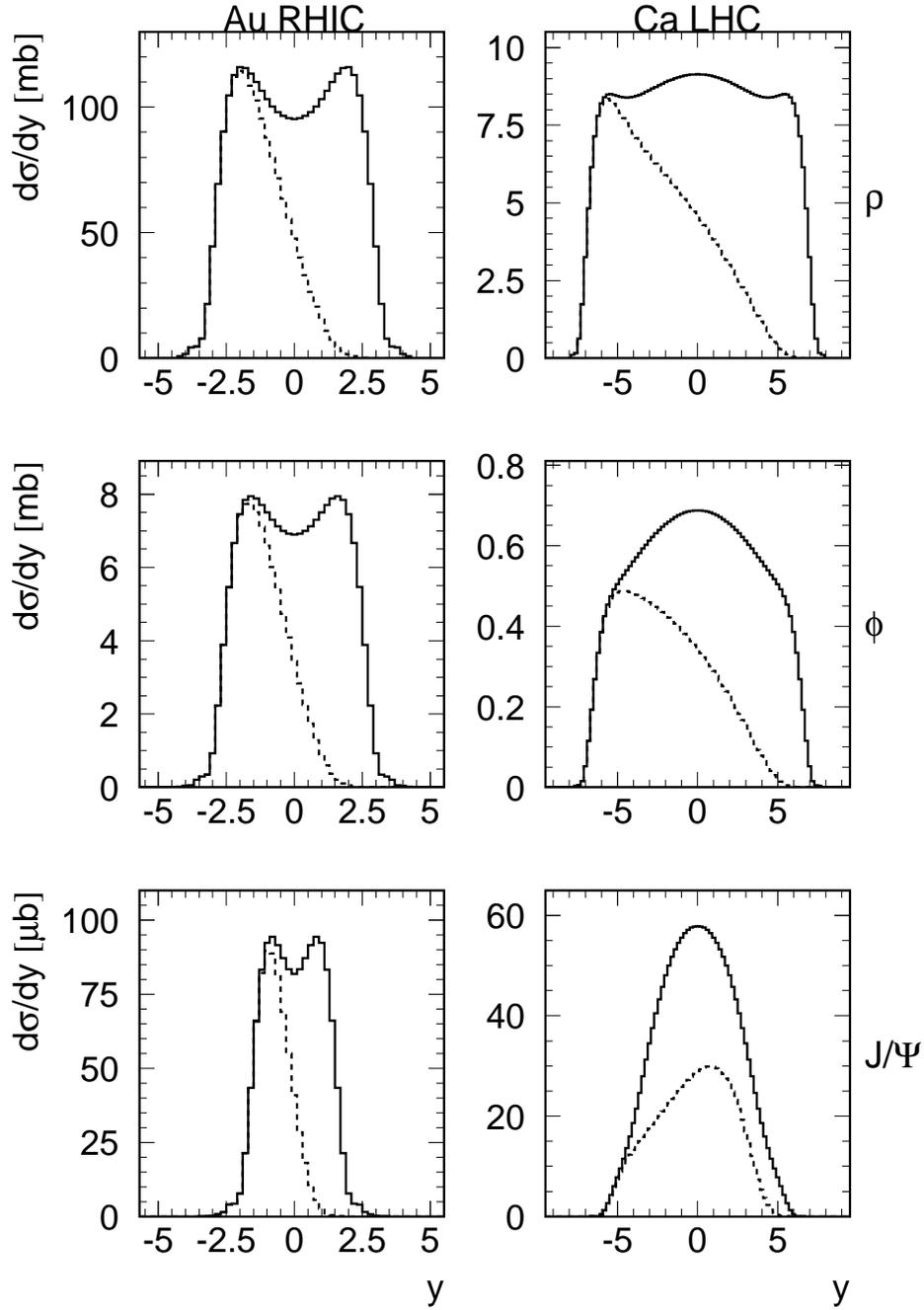}}
\vskip .2 in
\caption[]{The rapidity distribution of produced $\rho$ (top),
$\phi$ (center) and J/$\psi$ (bottom).  The left panels show
are for gold beams at RHIC, while the right panels are for
calcium beams at LHC.  The solid line is the total
production, while the dashed line is for photons coming
from the nucleus on the left.}
\label{dndy}
\end{figure}

The final state rapidity is determined by the ratio of the
photon energy and $t_{||}$ in the lab frame:
\begin{equation}
y= {1\over 2} \ln {k\over \sqrt{|t_{min}|}} = \ln {k\over 2M_V}
\end{equation}
so $d\sigma/dy=kd\sigma/dk$ and $d\sigma/dy$ is found by a change of
variables.  For the $\rho$, the cross section must be integrated over
$M_{\pi\pi}$.  $dN/dy$ is shown in Fig.~\ref{dndy}, for the $\rho$
(Breit-Wigner), $\phi$ and $J/\psi$, for production with gold at RHIC
and calcium at LHC. The cross section is largest at low photon
energies, corresponding to $y<0$, because the photon flux drops as $k$
rises. The peaking is larger for the $\rho$ and $\omega$, where meson
exchange increases the cross section at low energies.

At RHIC, when photons from both nuclei are added together, separate
peaks appear in $dN/dy$. At LHC, the higher energies spread the
distribution over a much wider rapidity range, and the double peaked
structure largely disappears, except for the $\rho$.

The total perpendicular momentum spectrum is dominated by the two
nuclear form factors, through the cutoff on the emitted photon
$p_\perp$ and through the slope $b$ of $d\sigma/dt$.  The (quadrature)
sum of these is roughly $\sqrt{2}\hbar/R_A$, or about 45 MeV/c for
gold.  This cutoff is a clear signature of coherent interactions,
shared with $\gamma\gamma$ and coherent double Pomeron interactions.
Because of the exclusive production and small $p_\perp$, these events
can be easily separated from background such as grazing nuclear
collisions, beam gas interactions and incoherent photonuclear
interactions\cite{hadron97}.

Other authors have considered coherent double-Pomeron interactions in
heavy ion collisions\cite{engel}.  If reactions where the nuclei
collide are excluded, the cross section is very small, and depends
critically on the range of the Pomeron\cite{muller}\cite{schram}.

\section{Multiple Vector Meson Production}

\begin{figure}
\setlength{\epsfxsize=0.77\textwidth}
\setlength{\epsfysize=0.43\textheight}
\centerline{\epsffile{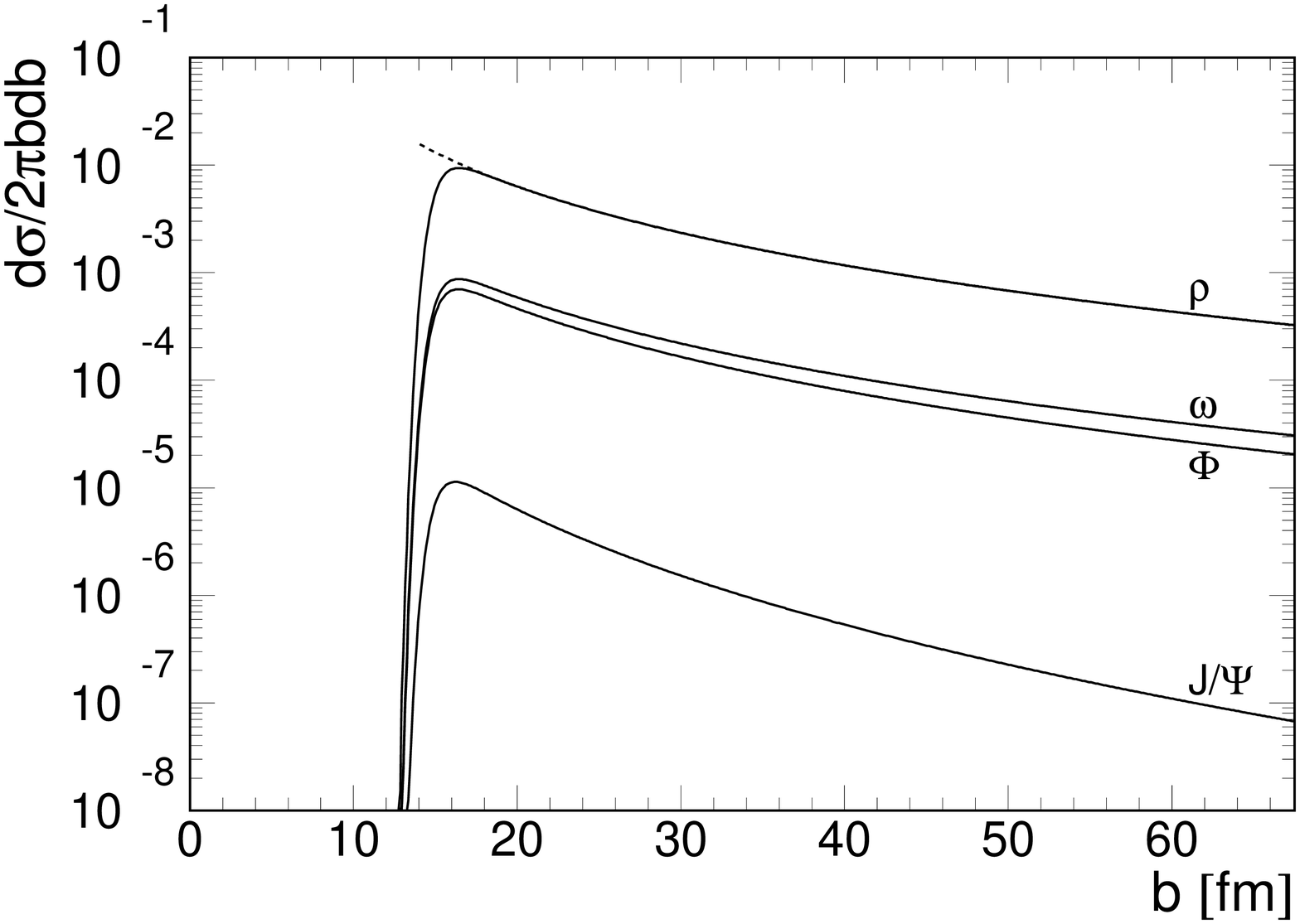}}
\setlength{\epsfxsize=0.77\textwidth}
\setlength{\epsfysize=0.43\textheight}
\centerline{\epsffile{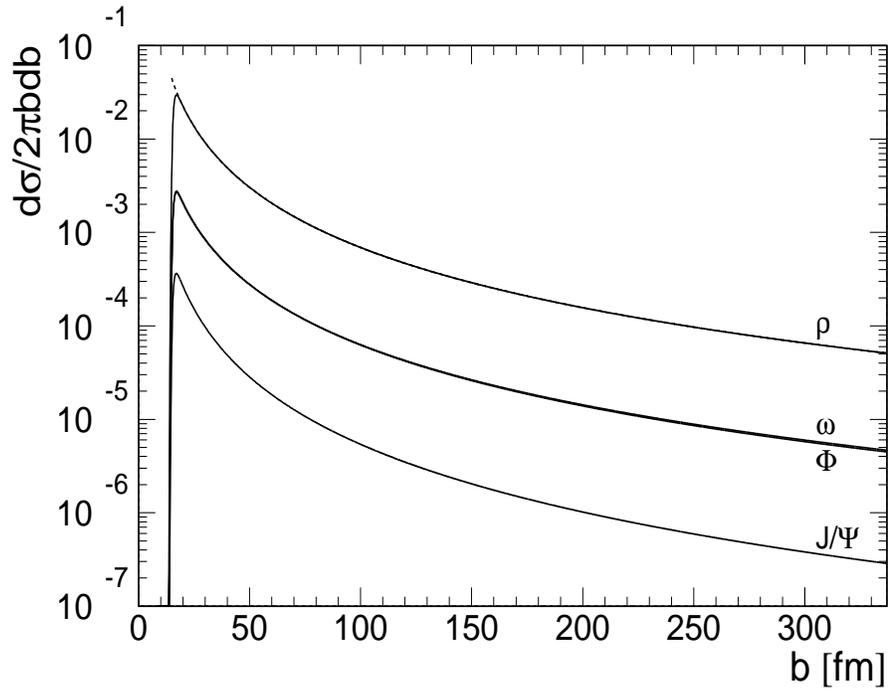}}
\caption[]{Probability of vector meson production as a function of
collision impact parameter for (a) gold at RHIC and (b) lead at LHC.
The solid lines show the Woods-Saxon hadronic interaction probability,
while the dashed lines shows the result of a hard sphere calculation
for the $\rho$.}
\label{probability}
\end{figure}

Because the cross sections are so large, the probability of having
multiple interactions in a single nucleus-nucleus encounter is
nonnegligible.  This can be quantified by considering the meson
production probability $P(V)=d\sigma/2\pi bdb$ for a single collision.
Figure~\ref{probability} shows this probability for various mesons for
(a) gold beams at RHIC and (b) lead beams at LHC; the probabilities
reach the 1\% level.  For the $\rho$, we show results from two
different hadronic interactions models: the Woods-Saxon model,
Eq. (\ref{eqhint}), and also a hard sphere $b>2R_A$ result.  Although
the two models predict very similar overall rates, the probability is
significantly affected around $b=2R_A$, doubling the maximum
probability at LHC.

This probability is high enough that, even in the absence of
correlations\cite{future}, significant numbers of vector meson pairs
should be produced.  Neglecting correlations, the pair production
probability is $P(V)^2/2$ for identical pairs and $P(V_1)P(V_2)$ for
nonidentical pairs.  These cross section and corresponding yearly
rates are tabulated in Table~\ref{sigmapair}.

Triples and higher multiples are also possible.  Of course, final
state interactions may affect what is observed in a detector.
 
\section{Interplay between photonuclear and two-photon interactions}

Two-photon physics is expected to be a significant activity at RHIC
\cite{usRHIC}\cite{goldflash} and LHC\cite{baurrev}\cite{FELIX}.
Because the kinematics for both photonuclear interactions and
two-photon collisions is dominated by the coherent coupling to the two
nuclei, and consequently by the nuclear form factors, the reactions
have similar kinematics.  So, the specific pathway may not be
determinable from the final state.

The rates for photonuclear interactions are considerably higher than
for comparable two-photon reactions. For example, the $\phi$
production rate with gold beams at RHIC is about 8 Hz.  For
comparison, the rates for $\gamma\gamma\rightarrow \eta'$ and
$\gamma\gamma\rightarrow f_0(980)$\cite{sn347} are 0.13 and 0.02 Hz
respectively.  At higher masses, the situation is similar, and
photonuclear vector meson production may overshadow two-photon
production of scalar and tensor final states, complicating meson
spectroscopy.  It is worth noting that the photonuclear $\rho\rho$
rate is many times larger than that expected from $\gamma\gamma$
production\cite{FELIX}.  On the other hand, the same techniques
developed to study $\gamma\gamma$ physics at heavy ion colliders are
effective at selecting photonuclear interactions.

The similarity of the two interactions can lead to some interesting
opportunities.  For example, when the final states are
indistinguishable, interference can occur between the photonuclear and
$\gamma\gamma$ production channels.  Interference between leptonic
decays of photonuclear produced vector mesons and Bethe-Heitler
$e^+e^-$ pairs has been studied previously\cite{leith}; at future
heavy ion colliders, many more channels will be accessible.

\begin{figure}
\setlength{\epsfxsize=0.7\textwidth}
\setlength{\epsfysize=0.4\textheight}
\centerline{\epsffile{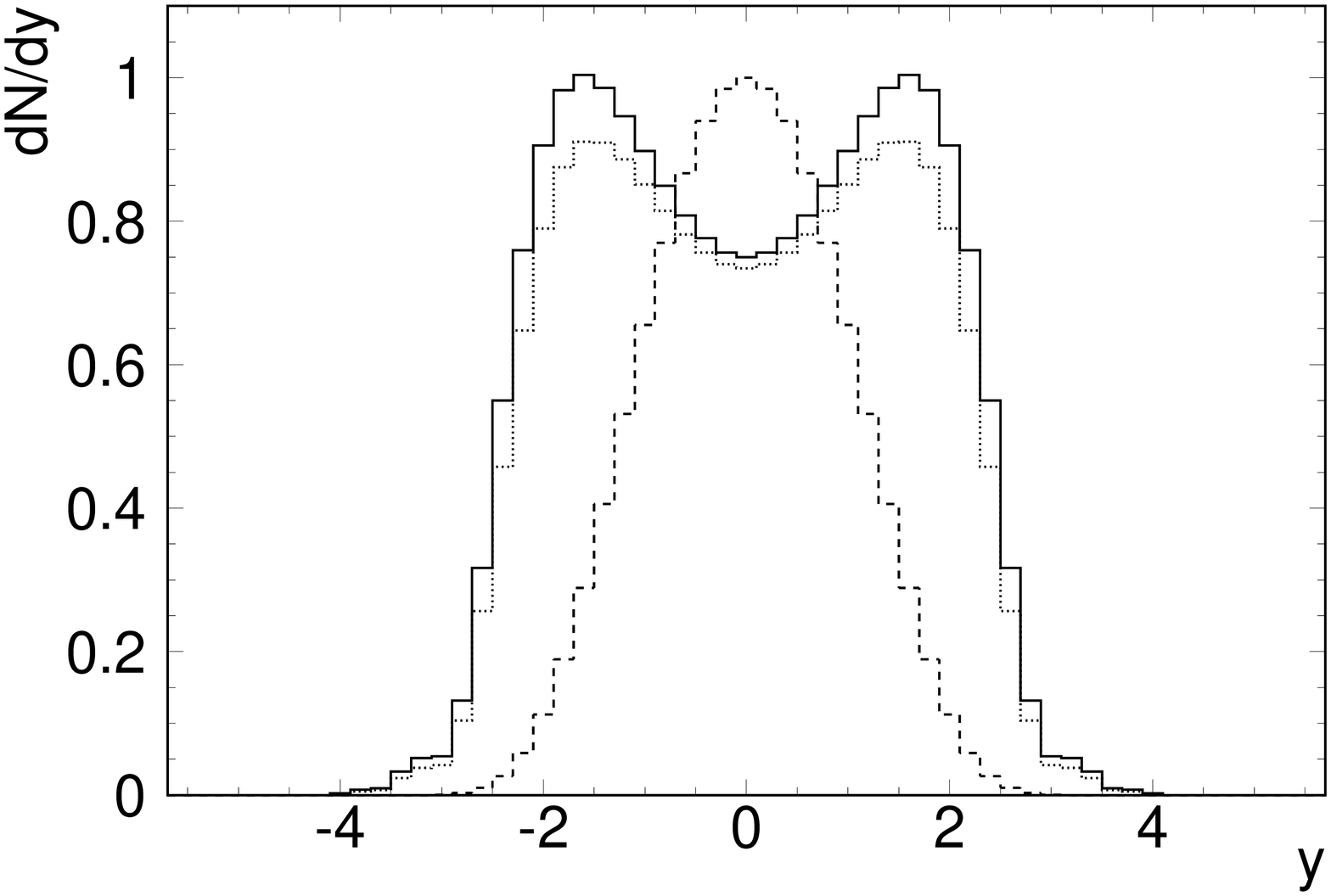}}
\label{dndycompare}
\caption[]{Comparison of the rapidity distributions for two-photon
production of the $f_2(1270)$ (dashed histogram) and two models of
photonuclear production of a hypothetical $X(1270)$, for gold beams at
RHIC.  The solid histogram is for an $X(1270)$ with couplings matching
the $\omega$, while the dotted histogram corresponds to Pomeron
coupling only.}
\label{frapcomp}
\end{figure}

Despite the similarity, the different $dN/dy$ spectra may allow
statistical separation of $\gamma\gamma$ and photonuclear
interactions.  This is because the photon flux scales as $1/k$, while
the Pomeron/meson couplings have a much weaker energy dependence.  The
$dN/dy$ distribution for photonuclear production is broader than for
two-photon production.  This is especially true for photon-meson
interactions, where the $1/k$ photon spectrum combines with a cross
section that drops with energy to form two peaks at large positive and
negative rapidity, corresponding to low energy photons from the two
nuclei in turn.  Figure \ref{dndycompare} compares the two
distributions for $\gamma\gamma\rightarrow f_2(1270)$ and for
production of a hypothetical vector meson with the same mass.  Two
vector meson models are shown: one based on the $\omega$ couplings and
another with only Pomeron coupling.  The two are very similar at
$y=0$, but the $\omega$ model peaks about 10\% higher.

It may also be possible to separate the two classes by their different
impact parameter dependence. A two-photon interaction can occur at a
significant distance from both nuclei, while a photonuclear
interaction must occur inside or very near a nucleus.  So, the two
classes will have a different impact parameter dependence.  In a
collision, photon exchange can leave the nuclei excited in a giant
dipole resonance, regardless of other activity; the excitation
probability depends on the impact parameter\cite{baurrev}.  This may
allow us to statistically separate the two classes.

In a broader study, it will also be possible to separate the two
classes by measuring how the rates vary with $A$; the meson and
Pomeron flux rises should be less $A$ dependent than photons.  This
study will require data with a variety of different beam species.

\section{Conclusions}

We have calculated the rates and rapidity distribution for exclusive
photonuclear production of vector mesons in heavy ion collisions,
using a Glauber model calculation.  The LHC results involve photons
with energies considerably above those currently available.  Although
this adds to the physics interest, it also introduces some uncertainty
into the rates presented here.  For the $J/\psi$, in particular, the
spectrum must soften at some photon energy, and the rates given here
may be overestimates.

The production rates are large enough that heavy ion colliders could
be used as vector meson factories.  The $\phi$ and $J/\psi$ production
rates at LHC are comparable to those at existing or planned meson
factories based on $e^+e^-$ annihilation.

These rates are higher than comparable two-photon or double-Pomeron
interactions.  Thus, these photonuclear interactions should be the
dominant source of exclusive interactions at heavy ion colliders.  In
addition to single interactions, multiple vector meson production from
the same ion pair should be measurable.  

We thank J\o rgen Randrup for suggesting the hard sphere plus Yukawa
potential form factor.  We would like to acknowledge useful
conversations with Stan Brodsky and Ralph Engel.  This work was
supported by the US DOE, under contract DE-AC-03-76SF00098.

\begin{table}
\caption[]{Luminosity and beam kinetic energy for several heavy ion
beams at RHIC and LHC.  Because RHIC will be dedicated to heavy ion
acceleration, it is likely to run a wider variety of beams than
LHC. The RHIC luminosities are from Ref.~\cite{RHIC}.  Different
references quote somewhat different heavy ion luminosity for LHC;
these numbers are calculated from Table 3 of Ref. \cite{LHClum},
assuming 2 experiments, and that (as suggested) average luminosity is
45\% of peak luminosity.}
\begin{tabular}{lrrr}
Particle	& Machine& Max Beam Energy & Design Luminosity \\
\hline
Gold		& RHIC	& 100 GeV/n	& $2\times10^{26}cm^{-2}s^{-1}$	\\
Iodine		& RHIC	& 104 GeV/n	& $2.7\times10^{27}cm^{-2}s^{-1}$\\
Silicon		& RHIC	& 125 GeV/n	& $4.4\times10^{28}cm^{-2}s^{-1}$\\
Lead		& LHC	& 2.76 TeV/n	& $1\times10^{26}cm^{-2}s^{-1}$	\\
Calcium		& LHC	& 3.5 TeV/n	& $2\times10^{30}cm^{-2}s^{-1}$	\\
\end{tabular}
\label{RHIClum}
\end{table}

\begin{table}
\caption[]{Constants for $\gamma p\rightarrow Vp$ production.  The
slopes $b_V$ are in GeV$^{-2}$, while $X$ and $Y$ are in $\mu$barns,
for $W$ in GeV\cite{sigmacon}.  The $f_v$ values are from
Eq. (\ref{fvs}).}
\begin{tabular}{lrrrrrr}
Meson & $b_V$ & $X$ & $\epsilon$ & $Y$ & $\eta $ & $f_v^2/4\pi$ \\
\hline
$\rho^0$ & 11 & 5.0  & 0.22 & 26.0 & 1.23 & 2.02 \\
$\omega$ & 10 & 0.55 & 0.22 & 18.0 & 1.92 & 23.1\\
$\phi$   & 7 & 0.34 & 0.22 &      &       & 13.7\\
$J/\psi$ & 4 & 0.0015 & 0.80 &    &       & 10.4\\
\end{tabular}
\label{HERAsigma}
\end{table}

\begin{table}
\caption{Cross sections, in millibarns, for production of vector
mesons for the beams given in Table 2.}
\begin{tabular}{lrrrrr}
Meson 	&  RHIC-Au	& RHIC-I & RHIC-Si & LHC-Pb & LHC-Ca 	\\
\hline
$\rho^0$& 590		& 230	 & 8.4	& 5200	& 120	\\
$\omega$& 59		& 24	 & 0.9	& 490	& 12	\\
$\phi$	& 39		& 14	 & 0.4	& 460	& 7.6	\\
J/$\psi$& 0.29		& 0.11	 & 0.0036& 32	& 0.39	\\
\end{tabular}
\label{sigmavm}
\end{table}

\begin{table}
\caption{Meson production rates, in Hz, at design luminosity for
various beams.}
\begin{tabular}{lrrrrr}
Meson 	&  (RHIC-Au)	& RHIC-I & RHIC-Si & LHC-Pb & LHC-Ca 	\\
\hline
$\rho^0$& 120		& 620	& 370	& 520	& 230,000 \\
$\omega$& 12		& 63	& 42	& 49	& 23,000	\\
$\phi$	& 7.9		& 39	& 18	& 46	& 15,000	\\
J/$\psi$&  0.058 	& 0.30  & 0.16  & 3.2 	& 780  \\
\end{tabular}
\label{ratevm}
\end{table}

\begin{table}
\caption{Cross sections and rates for production of vector mesons
pairs in gold beams at RHIC and lead beams at LHC.  The yearly rates
are for $10^{7}$ second/year at RHIC and $10^{6}$ second/year of
operation at LHC.  The LHC $\rho J/\psi$ rates should be used with
caution because of the energy extrapolation.}
\begin{tabular}{lrrrrr}
Meson 	&  $\sigma$($\mu$b) (RHIC-Au)	& Rate/year (RHIC-Au) &
 $\sigma$($\mu$b) (LHC-Pb) & Rate/year (LHC-Pb) 	\\
\hline
$\rho^0\rho^0$& 720	& 1,400,000 	& 8,800	& 880,000 \\
$\omega\omega$	& 6.2	& 12,000 	& 73	& 7,300   \\
$\phi\phi    $	& 3.8	& 7,500		& 76	& 7,600   \\
$\rho^0\omega$	& 133	& 270,000	& 1,600	& 160,000 \\
$\rho^0\phi  $	& 104	& 210,000 	& 1,640 & 160,000 \\
$\omega\phi  $	& 9.6	& 19,000	& 150	& 15,000  \\
$\rho^0J/\psi$  & 1.3 	& 2,700		& 200	& 20,000	\\
\end{tabular}
\label{sigmapair}
\end{table}

\vfill\eject

\vfill\eject

\vfill\eject

\vfill\eject

\vfill\eject

\end{document}